\documentclass[%
reprint,
 amsmath,amssymb,
 aps,
]
{revtex4-1}

\usepackage{graphicx}
\usepackage{dcolumn}
\usepackage{bm}
\usepackage{subfigure}
\usepackage{mathtools,slashed}
\usepackage{amsmath,amsfonts,amsthm}
\usepackage{tikz}       
\usepackage{tikz-feynman}

\begin{document}


\title{Photon time delay resulting from vacuum fluctuations:\\ the vacuum as a dielectric }

\author{G. B. Mainland}%
 \email{mainland.1@osu.edu}
\affiliation{Department of Physics, The Ohio State University at Newark, 1179 University Dr., Newark, OH 43055}

\author{Bernard Mulligan}%
 \email{mulligan.3@osu.edu}
\affiliation{Department of Physics, The Ohio State University, Columbus, OH 43210}


\author{}
\affiliation{}

\setcounter{equation}{0}

\date{\today}

\begin{abstract}
It is not possible to detect  a vacuum fluctuation without a test particle interacting with the vacuum fluctuation in a measurable manner.  In the quantum electrodynamics calculation presented here, a  photon traveling through the vacuum is used as the test particle to determine properties of the vacuum resulting from vacuum fluctuations. In particular, this article  (1) discusses the mathematical  procedure for describing a lepton-antilepton bound state that is a vacuum fluctuation, (2) describes the photon interacting with and being incorporated into the bound state to form a quasi-stationary bound state, and (3) calculates the electromagnetic decay rate of the quasi-stationary bound state.
 
\end{abstract}


\maketitle 

\section{Introduction}

According to perturbation theory in quantum electrodynamics, as a photon travels through the vacuum it can be viewed as splitting into a charged, virtual, particle-antiparticle pair that then annihilates, emitting a photon identical to the original.  Such interactions contribute to the photon propagator but, even to infinite order, do not change the location of the pole of the propagator.  Because the position of pole does not change, the speed of a photon does not change as a result of interacting with virtual particles. In this article the term ``virtual particle''  refers exclusively to a particle associated with a Green's function or, equivalently, with a propagator that arises in perturbative calculations using  quantum field theories. 

In addition to virtual particles, there is a second concept associated with the vacuum, that of ``vacuum fluctuations'' (VFs).  The term VFs  refers exclusively to  particles that appear spontaneously within the vacuum and are not associated with a Green's function.  VFs cannot be observed directly but, as will be demonstrated in the present article, can be observed indirectly by the effect of their interactions with photons moving through the vacuum. Using only quantum electrodynamics to describe a photon traveling through the vacuum, it is possible to show that the presence of certain types of VFs slows the progress of photons.

The result presented here can be viewed as being foreshadowed by Dimopoulos, Raby, and Wilczek who state,``The vacuum is a dielectric.''\cite{Dimopolos:91}  Their statement is made in reference to the well-understood electromagnetic, weak, and color charge renormalization.  Here, however,  the possibility is examined that the statement, ``The vacuum is a dielectric,'' is more generally true. Specifically, the energy  associated with a particular type of VF, a lepton-antilepton pair,  is postulated to increase until the lepton and antilepton are on mass shell, so that they can be treated as external particles, and are bound into an atom in its ground state. Not only does the binding into atoms minimize the transient violation of conservation of energy allowed by what is commonly referred to as the time-energy uncertainty principle, it also provides atoms that can interact with photons. The term ``dielectric'' can then be used in the usual sense:  a photon passing through a dielectric is slowed by its interactions with atoms in the dielectric, a concept familiar from discussions of a physical dielectric \cite{Kramers:24,Rossi:59, Feynman:63}.  Because there is a nonzero, finite lifetime associated with this quasi-stationary state, there will be a delay in the progress of the photon through the vacuum, this delay being increased with each interaction of the photon with a lepton-antilepton VF.  The effect of this interaction will appear in the propagator only as a change in the value of $c$, not as a change in the position of the pole.  As required by conservation of energy, when the quasi-stationary state annihilates, the vacuum reabsorbs the released energy.

\section{Calculation of the decay rate of photon-excited parapositronium}

Here the decay rate is calculated for electron-positron pairs that are VFs  and appear as parapositronium, the positronium state with the lowest energy that also happens to have zero angular momentum.   The formula for the decay rate immediately generalizes to yield decay rates for VFs of  muon-antimuon and tau-antitau pairs that are bound into atoms in their ground states.   These three decay rates play an essential role in determining the speed of photons as they propagate through the vacuum.   The decay is kinematically forbidden for ordinary parapositronium but is allowed for  parapositronium that is a VF since the latter parapositronium does not enter into overall energy conservation because it contributes no net energy.

Labeling the initial (incident) and final (emitted) photons, respectively, by $\gamma_i$ and $\gamma_f$, to lowest order the two Feynman diagrams that contribute to the process $\gamma_i$+positronium  that is a VF $\rightarrow \gamma_f$ are shown in FIG.~\ref{fig:1}.  \footnote{The cross section is calculated for excited  positronium that is a VF, and the restriction to parapositronium is not made until the decay rate is calculated.}
\setcounter{figure}{0}
\begin{figure}[h]
\begin{tikzpicture}
\begin{feynman}
\vertex (d);
\vertex [right=of d] (df);
\vertex [right=of df] (f){\(\gamma_{f}\)}; 
\vertex [below left=of d] (de);
\vertex [below left=of de] (e);
\vertex [left=of e] (ec);
\vertex [left=of ec] (c){\(\gamma_{i}\)};
\vertex [above left=of e] (eb); 
\vertex [above left=of eb] (b){\(e^{+}\)};
\vertex [above left=of d](da);
\vertex [above left=of da](a){\(e^{-}\)};
\diagram* {
(a)  --[fermion, edge label=\(p_{-}\)]  (d) -- [fermion, edge label=\(-p_{+}-k_i\)] (e)-- [fermion, edge label=\(-p_{+}\)] (b), 
(d) -- [photon, momentum'=\(k_f\)] (f),
(c) -- [photon, momentum'=\(k_i\)] (e),
}; 
\end{feynman}
\end{tikzpicture}
\begin{center}
(a)
\end{center}

\begin{tikzpicture}
\begin{feynman}

\vertex (d);
\vertex [left=of d] (db);
\vertex [left=of db] (b){\(\gamma_{i}\)}; 

\vertex [above left=of d] (da);
\vertex [above left=of da] (a){\(e^{-}\)};

\vertex [below left=of d] (de);
\vertex [below left=of de] (e);

\vertex [right=of e] (ef);
\vertex [right=of ef] (f){\(\gamma_{f}\)}; 

\vertex [above left=of e] (ec);
\vertex [above left=of ec] (c){\(e^{+}\)};

 \diagram* {
(a)  --[fermion, edge label=\(p_{-}\)]  (d) -- [fermion, edge label=\(p_{-}+k_i\)] (e)-- [fermion, edge label=\(-p_{+}\)] (c), 
(b) -- [photon, momentum'=\(k_i\)] (d),
(e) -- [photon, momentum'=\(k_f\)] (f),
}; 

\end{feynman}
\end{tikzpicture}
\begin{center}
(b)
\end{center}

\caption{ (a) Photon $\gamma_i$ interacts with a positron that then annihilates with an electron, emitting  photon $\gamma_f$. (b) Photon  $\gamma_i$ interacts with an electron that then annihilates with a positron, emitting  photon $\gamma_f$.}
\label{fig:1}
\end{figure}
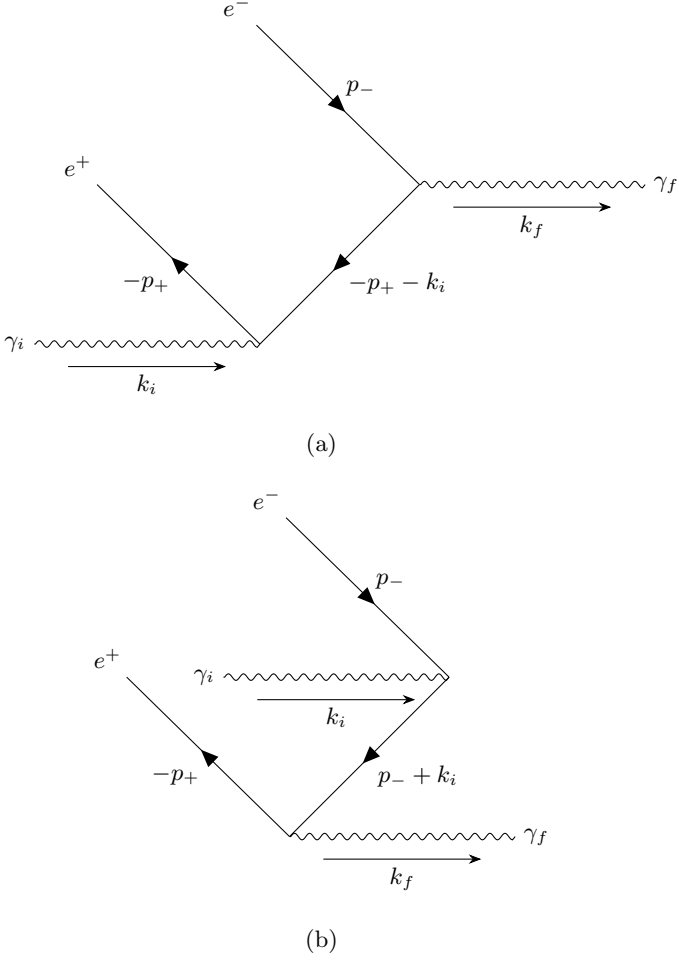

In the  diagrams $p_-, p_+, k_i,$ and $k_f$ are, respectively, the four-momenta of the electron, positron, initial  photon, and final photon.

For ordinary positronium the process is kinematically forbidden.  In the center-of-mass rest frame of positronium, $\mathbf{p_-}+\mathbf{p_+}=0$. Therefore, in this frame \footnote{Theorists' units are used so $\hbar$ and $c$ are each replaced by 1.},
\begin{subequations}\label{eqn:2}
\begin{align}
\label{eqn:2a}
p_-&=(E_-, \mathbf{p_-})=(\sqrt{m_e^2+\mathbf{p_-}^2},\mathbf{p_-})\,, \\
\label{eqn:2b}
p_+&=(E_+, \mathbf{p_+})=(\sqrt{m_e^2+\mathbf{p_+}^2},\mathbf{p_+})\,,\nonumber\\
&=(\sqrt{m_e^2+\mathbf{p_-}^2},-\mathbf{p_-})=(E_-, -\mathbf{p_-})\,,\\
\label{eqn:2c}
k_i&=(\omega_i,\mathbf{k_i})=(|\mathbf{k_i}|,\mathbf{k_i})\,,  \\
\label{eqn:2d}
k_f&=(\omega_f,\mathbf{k_f})=(|\mathbf{k_f}|,\mathbf{k_f})\,. 
\end{align}
\end{subequations}
Conservation of energy and momentum requires $p_++p_-+k_i=k_f$.  Squaring both sides of the above equation yields
\begin{equation}\label{eqn:3}
E_-^2+E_-\omega_i=0\,.
\end{equation}
Eq.  \eqref{eqn:3} cannot be satisfied for ordinary positronium since both terms on the left-hand side are positive. However, after a photon excites   positronuim that is a VF, a photon can be emitted, but only when the positronium vanishes into the vacuum because only then does $E_-\rightarrow 0$, allowing  \eqref{eqn:3} to be satisfied.

When performing an electrodynamics calculation, if a factor exp$\pm (ip\cdot x)$ is associated with a particle that is a VF when it appears in its initial state, a factor  exp$\mp (ip\cdot x)$ is associated  with a particle that is a VF  when it vanishes into the Grid.  This just eliminates the contribution of the particle that is a VF to overall energy-momentum conservation. When progressing along an energy-momentum line in a Feynman diagram, the energy-momentum associated with a particle that is a VF is not further used after the particle vanishes.  Since the  electron and positron that are VFs are treated as external particles,  they must be on-shell.  Using the notation  of \cite{Bjorken:64}, the S-matrix for the transition ``photon +   positronium that is a  VF $\rightarrow$ photon'' is 
\begin{align}\label{eqn:4}
S_{\rm fi}=& \frac{e^2}{V^2}\sqrt{\frac{m_e}{E_+}}\sqrt{\frac{m_e}{E_ -}}\frac{1}{\sqrt{2\omega_i}}\frac{1}{\sqrt{2 \omega_f}}\,(2\pi)^4\delta(k_i-k_f) \times \nonumber \\
&\bar{v}(p_+,s_+)[(-i\slashed{\epsilon}_i)\frac{i}{-\slashed{p}_+ -\slashed{k}_i-m_e} (-i\slashed{\epsilon}_f)+\nonumber\\
&\hspace{1.5 cm}(-i\slashed{\epsilon}_f)\frac{i}{\slashed{p}_- +\slashed{k}_i-m_e} (-i\slashed{\epsilon}_i)]u(p_-,s_-)\,.
\end{align}
In \eqref{eqn:4} the fermion wave functions are normalized to unit probability in a box of volume $V$.  The equality $1/(\slashed{p}\pm m_e)=(\slashed{p}\mp m_e)/(p^2-m_e^2)$ is used to rewrite the above two propagators. Then the equation
\begin{equation}\label{eqn:5}
\slashed{a} \slashed{b}=-\slashed{b} \slashed{a}+2a\cdot b I\,,
\end{equation}
where $a$ and $b$ are four-vectors and $I$ is the identity matrix, yields $\slashed{\epsilon}_i \slashed{p}_+=-\slashed{p}_+ \slashed{\epsilon}_i+2p_+\cdot \epsilon_i I$ and 
$\slashed{p}_- \slashed{\epsilon}_i =-\slashed{\epsilon}_i\slashed{p}_-+2p_-\cdot \epsilon_i I$. Also $\bar{v}(p_+,s_+)\slashed{p}_+=-\bar{v}(p_+,s_+)\,m_e$ and $\slashed{p}_-u(p_-,s_-)=m_e\,u(p_-,s_-)$, with the result that \eqref{eqn:4} can be rewritten as
\begin{align}\label{eqn:6}
S_{\rm fi}=&-i \frac{e^2}{V^2}\sqrt{\frac{m_e}{E_+}}\sqrt{\frac{m_e}{E_ -}}\frac{1}{\sqrt{2\omega_i}}\frac{1}{\sqrt{2 \omega_f}}\,(2\pi)^4\delta(k_i-k_f) \times \nonumber \\
&\bar{v}(p_+,s_+)\left[-\frac{ (2\epsilon_i\cdot p_++\slashed{\epsilon}_i\slashed{k}_i) \slashed{\epsilon}_f}{(p_++k_i)^2-m_e^2}+\right.\nonumber\\ 
&\hspace{2.0 cm}\left. \frac{\slashed{\epsilon}_f(2p_-\cdot\epsilon_i+\slashed{k}_i\slashed{\epsilon}_i)}{(p_-+k_i)^2-m_e^2}\right]u(p_-,s_-)\,.
\end{align}

To obtain the decay rate to lowest order, in $S_{fi}$ the  respective velocities $\mathbf{v_-}$ and $\mathbf{v_+}$  of the electron and positron can be neglected. Thus 
\begin{subequations}\label{eqn:7}
\begin{align}\label{eqn:7a}
&E_- \rightarrow m_e\,, \hspace{1.0 cm} E_+ \rightarrow m_e\,,\\
\label{eqn:7b}
 &p_\pm \rightarrow (m_e, \mathbf{0})\, .
 \end{align}
\end{subequations}
The respective polarization vectors $\epsilon_i$ and $\epsilon_f$ of the initial and final photons are chosen to be space-like: $\epsilon_i=(0,\boldsymbol{\epsilon_i})$,  $\epsilon_f=(0,\boldsymbol{\epsilon_f})$ where
\begin{subequations}\label{eqn:8}
\begin{align}\label{eqn:8a}
&k_i \cdot \epsilon_i=-\mathbf{k_i}\cdot \boldsymbol{\epsilon_i}=0\,,\\
\label{eqn:8b}
&k_f\cdot \epsilon_f=-\mathbf{k_f}\cdot \boldsymbol{\epsilon_f}=0\,.
\end{align}
\end{subequations}
Using \eqref{eqn:7}, \eqref{eqn:8}, and $k_i\cdot k_i=0$,
\begin{subequations}\label{eqn:9}
\begin{align}\label{eqn:8a}
&(p_\pm+k_i)^2-m_e^2=2m_e\omega_i\,,\\
\label{eqn:8b}
&\epsilon_i \cdot p_\pm=0\,.
\end{align}
\end{subequations}

With the aid of \eqref{eqn:9}, \eqref{eqn:6} becomes
\begin{align}\label{eqn:10}
S_{\rm fi}=&-i \frac{e^2}{V^2}\frac{1}{\sqrt{2\omega_i}}\frac{1}{\sqrt{2 \omega_f}}\frac{1}{2m_e \omega_i}\,(2\pi)^4\delta^4(k_i-k_f) \times \nonumber \\
&\bar{v}(p_+,s_+)(- \slashed{\epsilon}_i\slashed{k}_i \slashed{\epsilon}_f
+\slashed{\epsilon}_f\slashed{k}_i\slashed{\epsilon}_i)u(p_-,s_-)\,.
\end{align}

Since $k_i=k_f$,  it follows from \eqref{eqn:5} and \eqref{eqn:8} that $\slashed{k_i}$ anti-commutes  with both $\slashed{\epsilon}_i$ and  $\slashed{\epsilon}_f$, allowing \eqref{eqn:10} to be rewritten as
\begin{align}\label{eqn:11}
S_{\rm fi}=&-i \frac{e^2}{V^2}\frac{1}{\sqrt{2\omega_i}}\frac{1}{\sqrt{2 \omega_f}}\frac{1}{2m_e \omega_i}\,(2\pi)^4\delta^4(k_i-k_f) \times \nonumber \\
&\bar{v}(p_+,s_+)(\slashed{\epsilon}_i \slashed{\epsilon}_f
-\slashed{\epsilon}_f\slashed{\epsilon}_i)\slashed{k}_iu(p_-,s_-)\,.
\end{align}
Then,
\begin{align}\label{eqn:12}
|S_{\rm fi}|^2=&\frac{e^4}{V^4}\frac{1}{2\omega_i}\frac{1}{2 \omega_f}\frac{1}{4m_e^2 \omega_i^2}VT\,(2\pi)^4\delta^4(k_i-k_f) \times \nonumber \\
&\bar{v}(p_+,s_+)(\slashed{\epsilon}_i \slashed{\epsilon}_f
-\slashed{\epsilon}_f\slashed{\epsilon}_i)\slashed{k}_i u(p_-,s_-)\times \nonumber\\
&\bar{u}(p_-,s_-)\slashed{k}_i(\slashed{\epsilon}_f \slashed{\epsilon}_i
-\slashed{\epsilon}_i\slashed{\epsilon}_f) v(p_+,s_+)\,.
\end{align}
In \eqref{eqn:12} the interaction is assumed to occur in the time interval $-T/2<t<T/2$.  

Summing over the electron and positron spins, which converts $|S_{\rm fi}|^2$ into a trace denoted by $Tr$,
\begin{align}\label{eqn:13}
&\sum_{s_\pm} |S_{\rm fi}|^2=\frac{e^4}{V^4}\frac{1}{2\omega_i}\frac{1}{2 \omega_f}\frac{1}{16m_e^4 \omega_i^2}VT\,(2\pi)^4\delta^4(k_i-k_f)\times \nonumber \\ 
&Tr[(\slashed{p}_+-m_e)(\slashed{\epsilon}_i \slashed{\epsilon}_f-\slashed{\epsilon}_f\slashed{\epsilon}_i) 
\slashed{k}_i (\slashed{p}_-+m_e)\slashed{k}_i(\slashed{\epsilon}_f \slashed{\epsilon}_i
-\slashed{\epsilon}_i\slashed{\epsilon}_f)]\,.
\end{align}
Using \eqref{eqn:5} to reverse the order of $\slashed{k}_i\slashed{p}_-$ and  using $\slashed{k}_i\slashed{k}_i=k_i\cdot k_i\,I=0$, the following term that appears in the second line above can be simplified:
\begin{equation}\label{eqn:14}
\slashed{k}_i (\slashed{p}_-+m_e)\slashed{k}_i=2m_e\omega_i\slashed{k}_i\,.
\end{equation}
As mentioned previously, $\slashed{k}_i$ anti-commutes with both $\slashed{\epsilon}_i$ and  $\slashed{\epsilon}_f$, with the result that \eqref{eqn:13} becomes
\begin{align}\label{eqn:15}
&\sum_{s_\pm}  |S_{\rm fi}|^2=\frac{e^4}{V^4}\frac{1}{2\omega_i}\frac{1}{2 \omega_f}\frac{1}{8m_e^3 \omega_i}VT\,(2\pi)^4\delta^4(k_i-k_f) \times \nonumber \\
&Tr[(\slashed{p}_+-m_e)(\slashed{\epsilon}_i \slashed{\epsilon}_i-\slashed{\epsilon}_f\slashed{\epsilon}_i)(\slashed{\epsilon}_f \slashed{\epsilon}_i
-\slashed{\epsilon}_i\slashed{\epsilon}_f)\slashed{k}_i]\,.
\end{align}
Eq. \eqref{eqn:5} is used as necessary to reverse the order of $\slashed{\epsilon}_i$ and $\slashed{\epsilon}_f$ so as to obtain terms of the form $\slashed{\epsilon}_i\slashed{\epsilon}_i=\epsilon_i \cdot \epsilon_i\,I=- \boldsymbol{\epsilon}_i \cdot \boldsymbol{\epsilon}_i\,I=-I$ and $\slashed{\epsilon}_f\slashed{\epsilon}_f=-I.$ All nonzero traces in \eqref{eqn:15}  are then either of the form $Tr(\slashed{a}\slashed{b})$ or $Tr(\slashed{a}\slashed{b}\slashed{c}\slashed{d}$) that are easily simplified, yielding the result
\begin{align}\label{eqn:16}
\sum_{s_\pm}  |S_{\rm fi}|^2=&\frac{2}{m_e^2}\frac{e^4}{V^4}\frac{1}{2\omega_i}\frac{1}{2 \omega_f}VT\,(2\pi)^4\delta^4(k_i-k_f) \times \nonumber \\
&[1-(\epsilon_i \cdot \epsilon_f)^2]\,.
\end{align}

The average cross section is now calculated for photon-excited positronium that is a VF to annihilate and emit a photon: $|S_{fi}|^2$ is divided by $VT$ to form a rate per unit volume, divided by the  electron-positron flux $|\mathbf{v}_+-\mathbf{v}_-|/V$, and divided by the number of target particles per unit volume $1/V$.  Averaging over the spins of the electron and positron ($\frac{1}{4}\sum_{s_\pm})$,  summing over the polarizations of the final photon ($\sum_{\epsilon_f})$, averaging over the polarizations of the initial photon ($\frac{1}{2}\sum_{\epsilon_i})$, summing over the number of states  of the final photon in the momentum interval ${\rm d}^3\mathbf{k}_f$ $(\int V {\rm d}^3\mathbf{k}_f/(2\pi)^3)$, and averaging over the number of states  of the initial photon in the momentum interval ${\rm d}^3\mathbf{k}_i$ $(\int V {\rm d}^3\mathbf{k}_i/(2\pi)^3)$, an expression for the cross section $\sigma$ is obtained:
\begin{align}\label{eqn:17}
\sigma=&\frac{1}{4}\sum_{s_\pm}\sum_{\epsilon_f}\frac{1}{2}\sum_{\epsilon_i}\int \frac{V{\rm d}^3\mathbf{k}_f}{(2\pi)^3} \int \frac{V{\rm d}^3\mathbf{k}_i}{(2\pi)^3}\times\nonumber\\
&\frac{|S_{fi}|^2}{VT}\frac{1}{\frac{|\mathbf{v}_+-\mathbf{v}_-|}{V}}\frac{1}{\frac{1}{V}} \,.
\end{align}
Using \eqref{eqn:16}, \eqref{eqn:17} becomes
\begin{align}\label{eqn:18}
\sigma=&\frac{\alpha^2}{m_e^2}\frac{1}{|\mathbf{v}_+-\mathbf{v}_-|}\sum_{\epsilon_f}\sum_{\epsilon_i}[1-(\epsilon_i \cdot \epsilon_f)^2]\times\nonumber\\
&\int_{-\infty}^\infty \frac{{\rm d}^3\mathbf{k}_f}{2\omega_f} \int_{-\infty}^\infty \frac{{\rm d}^3\mathbf{k}_i}{2\omega_i} \delta^4(k_i-k_f) \,.
\end{align}

Choosing the z-axis to point in the direction of $\mathbf{k_i}$, the unit polarization vectors for the initial photon  $\boldsymbol{\epsilon}_i^a$ and  $\boldsymbol{\epsilon}_i^b$ are chosen in the x- and y-direction, respectively.  Because the delta function in  \eqref{eqn:18} imposes the condition $k_f=k_i$, the unit polarization vectors for the final photon $\boldsymbol{\epsilon}_f^a$ and  $\boldsymbol{\epsilon}_f^b$ can also be chosen in the x- and y-direction, respectively. The sum over polarizations in \eqref{eqn:17} is  now easily performed:
\begin{subequations}\label{eqn:19}
\begin{align} \label{eqn:19a}
&\sum_{\epsilon_f}\sum_{\epsilon_i} 1=4\,,\\
\label{eqn:19b}
&\sum_{\epsilon_f}\sum_{\epsilon_i}(\epsilon_i \cdot \epsilon_f)^2=\nonumber\\
&(\epsilon_i^a \cdot \epsilon_f^a)^2+(\epsilon_i^b \cdot \epsilon_f^a)^2+(\epsilon_i^a \cdot \epsilon_f^b)^2+
(\epsilon_i^b \cdot \epsilon_f^b)^2\nonumber=\\
&(-1)^2+0+0+(-1)^2=2\,.
\end{align}
\end{subequations}

Using \eqref{eqn:2d} and the identity \cite{Schiff:55},
\begin{equation}\label{eqn:20}
\delta(\omega^2-a^2)=(1/2a)[\delta(\omega-a)+\delta(\omega+a)],\,a>0\,, 
\end{equation}
it is straightforward to show that
\begin{equation}\label{eqn:21}
\int_{-\infty}^\infty \frac{{\rm d}^3\mathbf{k}_i}{2\omega_i}=\int_{-\infty}^\infty {\rm d}^4k_i\,\delta(k_i^2) \theta(k_{i0})\,.
\end{equation}
The theta function $\theta(k_{i0})=0$ if $k_{i0}<0$ and $\theta(k_{i0})=1$ if $k_{i0}>0$. With the aid of \eqref{eqn:21}, the second line in \eqref{eqn:18}  can be rewritten as
\begin{align}\label{eqn:22}
&\int_{-\infty}^\infty \frac{{\rm d}^3\mathbf{k}_f}{2\omega_f} \int_{-\infty}^\infty \frac{{\rm d}^3\mathbf{k}_i}{2\omega_i}\, \delta^4(k_i-k_f)=\nonumber\\
&\int_{-\infty}^\infty \frac{{\rm d}^3\mathbf{k}_f}{2\omega_f}\,\delta(k_f^2)\,\theta(k_{f0})\,.
\end{align}
Factoring $k_f^2=k_{f0}^2-|\mathbf{k}_f|^2$ in the above $\delta$-function, using \eqref{eqn:20}, rewriting  ${\rm d}^3\mathbf{k}_f$ as ${\rm d}\Omega_f
|\mathbf{k}_f|^2{\rm d}|\mathbf{k}_f|$, performing the angular integration over ${\rm d}\Omega_f$, which yields a factor of $4\pi$, and then integrating over $|\mathbf{k}_f|$,
\begin{equation}\label{eqn:23}
\int_{-\infty}^\infty \frac{{\rm d}^3\mathbf{k}_f}{2\omega_f} \int_{-\infty}^\infty \frac{{\rm d}^3\mathbf{k}_i}{2\omega_i}\, \delta^4(k_i-k_f)=\pi\,.
\end{equation}

Substituting \eqref{eqn:19} and \eqref{eqn:23} into \eqref{eqn:18} yields the formula for the cross section for the annihilation into a photon of photon-excited positronium that is a VF,
\begin{equation}\label{eqn:24}
\sigma=\frac{2 \pi \alpha^2}{m_e^2}\frac{1}{|\mathbf{v}_+-\mathbf{v}_-|} \,.
\end{equation}

From the formula for the cross section, a formula for the decay rate is readily obtained. The logic is the same as that used to calculate the decay rate  for parapositronium decaying into two photons \cite{Wheeler:46}:  parapostronium, orthopositronium, and a photon have respective charge conjugation parities of +1, -1, and -1.  Thus photon-excited parapositronium has charge conjugation parity of -1 while photon-excited orthopositronium has charge conjugation parity of +1. Since electromagnetic interactions are invariant under charge conjugation,  photon-excited parapositronium, but not photon-excited orthopositronium, can decay into a single photon. 

In obtaining \eqref{eqn:24} the electron and positron spins were averaged over all four spins, resulting in the sum being divided by four. But the annihilating state is parapositronium, the singlet state.  Orthopositronium, the triplet state, does not contribute.  Since only one of the four spin states contributes to the cross section, the formula for the cross section should not have been divided by four, it should have been divided by the number one.  Thus the formula for $\sigma$ in \eqref{eqn:24} should be multiplied by a factor of four to obtain the cross section, abbreviated $\sigma_{\rm p-Ps}$, for the annihilation into a photon of  photon-excited parapositronium that is a VF,
\begin{equation}\label{eqn:25}
\sigma_{\rm p-Ps}=\frac{8 \pi \alpha^2}{m_e^2}\frac{1}{|\mathbf{v}_+-\mathbf{v}_-|} \,.
\end{equation}

For the annihilation of photon-excited parapositronium that is a VF into a photon, the electromagnetic  decay  rate $\Gamma_{p-Ps}$ is calculated using the mechanism for the annihilation of ordinary parapositronium \cite{Wheeler:46}.  The Schr\"odinger wave function $\psi(x)$ for parapositronium is just the ground-state hydrogen atom wave function with the reduced mass of hydrogen, which is approximately $m_e$, replaced by  $m_e/2$, the reduced mass of parapositronium:
\begin{equation}\label{eqn:26}
\psi(x)=\frac{1}{\sqrt{\pi}}\left(\frac{\alpha m_e}{2}\right)^{3/2}e^{- \alpha m_e\,r/2}\,.
\end{equation}
In the above formula $x$ is the magnitude of the vector $\mathbf{x}=\mathbf{x}_e-\mathbf{x}_p$ where $\mathbf{x}_e$ and $\mathbf{x}_p$ are, respectively,  the positions of the electron and the positron.  

The decay rate $\Gamma_{p-Ps}$ is the product of $\sigma_{\rm p-Ps}$ and the flux of a parapositronium atom, which is the relative velocity of approach of the electron and positron in parapositronium multiplied by $|\psi(0)|^2$, the probability density that the electron and positron collide and annihilate. 
\begin{align}\label{eqn:27}
\Gamma_{p-Ps}&=\sigma_{\rm p-Ps}|\mathbf{v}_+-\mathbf{v}_-|\,|\psi(0)|^2\,, \nonumber\\
&=\frac{8 \pi \alpha^2}{m_e^2}\frac{1}{|\mathbf{v}_+-\mathbf{v}_-|}|\mathbf{v}_+-\mathbf{v}_-|\,\frac{1}{\pi}\left(\frac{\alpha\,m_e}{2}\right)^3\,,\nonumber \\
&=\alpha^5m_e\,.
\end{align}
The above decay rate is twice that of ordinary parapositronium into two photons \cite{Wheeler:46}.  

From \eqref{eqn:27} it immediately follows that the  corresponding decay rates for  muon-antimuon and tau-antitau  bound, ground states that are VFs  are obtained by replacing  the  electron mass with the muon or tau mass, respectively.  As photons travel through the vacuum,  these three decay rates describe how photons interact with lepton-antilepton pairs that are VFs.

\section{Conclusion}

 When photons interact with virtual particles as described by quantum electrodynamics, the interactions do not affect the speed of the photons. In contrast, when photons interact with lepton-antilepton pairs that that are VFs, the speed of  photons is decreased similarly to the way that their speed is decreased in a dielectric consisting of ordinary matter. In an article to be submitted, the decay rates calculated here are used to determine how these interactions primarily determine the speed of photons in the vacuum.

\bibliography{Gamma}
\end{document}